\documentclass[aps,prd]{revtex4-2}
\usepackage{amsmath,amssymb,amsfonts,mathtools}
\usepackage{comment}
\usepackage{mathrsfs}
\usepackage{geometry}
\usepackage{enumitem}
\usepackage{booktabs}
\usepackage{xcolor}
\usepackage{physics}
\usepackage{hyperref}
\hypersetup{colorlinks=true,linkcolor=blue!60!black,citecolor=blue!60!black}

\newcommand{\re}{\operatorname{Re}}
\newcommand{\im}{\operatorname{Im}}
\begin{document}
\title{Holographic One-Point Functions and Geodesics in
\texorpdfstring{SdS$_3$}{SdS3}}
\author{Arundhati Goldar }
\email{d21086@students.iitmandi.ac.in}
\author{Nirmalya Kajuri}%
\email{nirmalya@iitmandi.ac.in}
 \author{Rhitaparna Pal}
 \email{d21089@students.iitmandi.ac.in}

\affiliation{%
School of Physical Sciences, IIT Mandi,\\ Himachal Pradesh 175005, India}%

\begin{abstract}
Grinberg and Maldacena showed that heavy thermal one-point functions in
AdS/CFT can encode complex geodesics reaching a  black hole singularity. We
study the de Sitter analogue in three-dimensional Schwarzschild--de Sitter
space, restricting to the finite cyclic quotients
$\mathrm{dS}_3/\mathbb Z_q$. We define a
one-point function through a differentiate dictionary and show that an
analogous result holds for the Bunch--Davies-prepared integral. Using the exact light field Green function, the bulk integral reduces exactly to a convergent one-dimensional transform, whose large-mass limit is controlled by the near-defect region and governed by a complex saddle. Its action reproduces a complex geodesic length from future infinity to the conical defect, whose real part is the renormalized boundary-to-horizon proper time and whose imaginary part is the horizon-to-defect distance. The imaginary part appears in the one-point coefficient normalized in the Lorentzian source basis; in the Euclidean basis the same saddle contributes only the real part, the two being related by a fixed connection factor.
\end{abstract}

\maketitle

\section{Introduction}

In the context of AdS/CFT
\cite{Maldacena:1997re,Gubser:1998bc,Witten:1998qj}, Grinberg and Maldacena
\cite{Grinberg:2020fdj} showed that the thermal one-point function of a
heavy boundary operator in the dual conformal field theory encodes the
complex geodesic length from the boundary insertion to the  black hole
singularity. Specifically, in the large-mass limit the one-point function
takes the schematic form
\begin{equation}\label{eq:ads_schematic}
 \langle \mathcal O\rangle_{\rm AdS}
 \propto
 \exp\!\left[-m\left(L_{\rm bd\to hor}
 +iL_{\rm hor\to sing}\right)\right].
\end{equation}
Here $L_{\rm bd\to hor}$ is the geodesic distance from the boundary to the
horizon and $L_{\rm hor\to sing}$ is the distance from the horizon to the
singularity; $m$ is the mass of the heavy probe.

This result is remarkable because the one-point function, a boundary
observable, encodes geometric information about the region behind the
horizon even though this region is classically inaccessible to a boundary
observer. Further studies of the connection between one-point functions and
bulk geodesics can be found in
\cite{Berenstein:2022nlj,David:2022nfn,David:2023uya,Singhi:2024sdr,
David:2024naf,Dey:2026ckh}.

We explore an extension of this result to three-dimensional
Schwarzschild--de Sitter space (SdS$_3$). These geometries do not possess a
curvature singularity but instead have a conical defect, which plays the role
of the singularity in the Grinberg--Maldacena setup. Unlike higher-dimensional
SdS geometries, SdS$_3$ offers two useful simplifications. First, it is
locally dS$_3$, which makes exact analytic results available. Second, it
possesses only a single cosmological horizon, further simplifying computations. We focus on a particular subclass of $SdS_3$---the finite
cyclic quotients
$\mathrm{dS}_3/\mathbb Z_q$, $q\geq2$, where
$\mathbb Z_q$ is the cyclic group of order $q$. The reason for this will be explained shortly.

Unlike AdS/CFT, dS/CFT
\cite{Strominger:2001pn,Witten:2001kn,Maldacena:2002vr,
McFadden:2009fg,Anninos:2011ui,Harlow:2011ke} is not an established
duality. Little is known about the putative dual theory, including where it
lives. We assume that it is located at the future boundary $\mathcal I^+$
and adopt the differentiate dictionary: the boundary
one-point function is obtained by differentiating $-i\log\Psi$ (where $\Psi$ is the
scalar wavefunctional) with respect
to the renormalized late-time source.

Following the analogous scalar construction in a BTZ background
\cite{Kraus:2016nwo,Grinberg:2020fdj}, we consider a heavy scalar $\varphi$
coupled to a light scalar $\chi$ through $g\varphi\chi^2$. In three
dimensions the Weyl tensor vanishes identically, so this interaction replaces
the curvature source used in the higher-dimensional Grinberg--Maldacena
setup. 

Let $P$ denote the future-boundary insertion. At leading order in $g$, the corresponding one-point
function has the form
\begin{equation}\label{eq:intro-bulk-integral}
 \langle\mathcal O(P)\rangle_K
 \;\propto\;
 g\int d^3x\,\sqrt{-g(x)}\,
 K(P;x)\,\langle\chi^2(x)\rangle_{\rm ren},
\end{equation}
where $K(P;x)$ is the heavy field bulk--boundary kernel and
$\langle\chi^2(x)\rangle_{\rm ren}$ is the renormalized coincidence limit
of the light field two-point function on the quotient.

The conical defect is a fixed locus of the quotient. Near this locus $\langle\chi^2(x)\rangle_{\rm ren}$ has the
expansion
\begin{equation}\label{eq:intro-near-defect-profile}
 \langle\chi^2\rangle_{\rm ren}
 =\frac{C_q}{\ell v}+O(v/\ell),
\end{equation}
where $\ell$ is the de Sitter radius, $v$ is the dimensionless transverse
radius, and $C_q$ depends only on the quotient. The $1/v$ pole is
universal---it arises from the short-distance Hadamard singularity of the
propagator.  

This
divergence compensates the small volume of the near-defect region. One then expects the large-mass behaviour of
\eqref{eq:intro-bulk-integral} to be governed by the defect. This heuristic expectation is borne out by exact computation.

The integral \eqref{eq:intro-bulk-integral} can be evaluated in closed
form. At the light-field mass $m_\chi^2\ell^2=1$, 
$\langle\chi^2\rangle_{\rm ren}$ on the quotient gives an elementary function of
the distance from the defect and the angular integrations can then be
carried out exactly. A single convergent integral remains, whose large-mass asymptotics are controlled by the defect.

The restriction to finite quotients, together with the choice of light-field mass, is what makes the above calculation
exact. Because the group is finite, the coincidence limit reduces to a sum of
$q-1$ image terms. Our choice of mass then makes each term is algebraic in
the distance from the defect. The singularities of the integrand form a
finite set of square-root loci that can be checked individually. Both these choices are
used in moving the integration contour onto a section where the integrand
is analytic, which allows us to complete the computation.
Neither restriction affects the leading order answer, however---the exponent below is independent of $q$
and of $m_\chi$

The answer depends on the choice of the quotient space kernel $K(P;x)$, or equivalently the covering space kernel (since the former is given by an image sum over the latter). The heavy field has two late-time falloffs, with the complex exponents
$1\pm i\nu$. Here \(\nu=\sqrt{M^2\ell^2-1}\) with where $M$ being the mass of the heavy field. Normalizing to a unit source means setting the coefficient of
one of these modes to one. As the the exponents are complex, the modes are
branched and that coefficient becomes definite only once the branch is
fixed. We use three source conventions for the covering space kernels, which we call bases. The Hartle--Hawking
convention fixes the branch on the regular Euclidean cap, where the
bulk--boundary invariant is positive, and continues from there. This gives
the kernel $K_{\rm HH}$ and the source coefficient $\mathcal J_{E,+}$. The
Lorentzian convention fixes it on the real section itself, through the
$+i0$ boundary value across the light cone. This gives the kernel that we denote by $K_L$ and the corresponding source
$\mathcal J_+$. Its conjugate uses the $-i0$ value together with the lower
preparation, giving $K_L^*$ and $\mathcal J_-$. The three differ
only by a branch factor. On the upper continuation the first two obey
\begin{equation}\label{eq:intro-kernel-relation}
 K_{\rm HH}=i e^{-\pi\nu/2}K_L.
\end{equation}
 The two conventions therefore assign different source coefficients to the
same bulk field configuration: expanding a given solution in both bases
gives $\mathcal J_+=i e^{-\pi\nu/2}\mathcal J_{E,+}$. Differentiating with
respect to the two coefficients then produces one-point functions
differing by the same factor. For instance, the one-point functions corresponding to $K_{\rm HH}$ and $K_L$ differ by a factor of $e^{-\pi\nu/2}$.

For the Lorentzian kernel $K_L$ defined above, we find
\begin{equation}\label{eq:main_result}
 \langle\mathcal O(P)\rangle_{K_L}
 \sim \mathcal P(\nu,P)e^{-iM L_+},
\end{equation}
while for its conjugate $K_L^*$:
\begin{equation}\label{eq:main_result}
 \langle\mathcal O(\theta_P)\rangle_{K_L}
 \sim \mathcal P(\nu,P)e^{-iM L_-}.
\end{equation}
$\mathcal P$ collects prefactors that do not affect the leading
heavy-mass exponential.
Here $L_\pm$ correspond to boundary-to-defect complex geodesic lengths. The geodesic calculation of Section~\ref{geodesic} returns both members of
this pair---extremizing the length over the complexified defect
worldline gives two nearest critical points related by reflection about
the real section. The corresponding complex geodesic lengths are $L_+$ and $L_-$.

The real part of $L_\pm$ is the renormalized boundary-to-horizon proper
time, while $|\im L_\pm|=\pi\ell/2$ is the proper distance from the
cosmological horizon to the defect. Thus the boundary-to-horizon segment
supplies the oscillatory phase, whereas the horizon-to-defect segment
supplies the exponential weight. Their roles are reversed from the AdS case. 

For $K_{\rm HH}$, the one point function behaves as $e^{-i\operatorname{Re}L_\pm}$, only the real part of $L_\pm$ features. Thus if one were to pick the Euclidean basis, the horizon-to-defect distance would not appear in the boundary one-point function. Whereas for the Lorentzian basis or its conjugate, it would. It is possible that the dual theory would single out one basis---we report the answers for all three.

The paper is organized as follows. Section~\ref{pre} sets up the
problem: the geometry of SdS$_{3}$ and its global coordinates, the
differentiate-dictionary definition of the one-point function, the three
source conventions and the relations between them, and the exact light-field
coincidence limit on the quotient. Section~\ref{op} specifies the
Bunch--Davies integration contour, reduces the vertex integral to a
one-dimensional radial transform, and extracts its heavy-mass asymptotics,
identifying the saddle selected by the prepared state.
Section~\ref{geodesic} computes the complex geodesic action
independently, fixes the branch of each of the two logarithms that enter, and
compares the resulting action values with the saddle exponent.
Section~\ref{disc} discusses what has been established, what depends
on the source convention, and directions for extending the result. Technical
material is collected in three appendices: Appendix~\ref{app1} constructs the
light-field Green function, Appendix~\ref{app2} proves the contour
deformation and the exact reduction, and Appendix~\ref{app3} rederives the
geodesic pair by continuation to $S^{3}$.

\section{One-point function for \texorpdfstring{\textsc{SdS}$_3$}{SdS3}: preliminaries
}\label{pre}

\subsection{Geometry of \texorpdfstring{\textsc{SdS}$_3$}{SdS3}}

A convenient form of the Lorentzian SdS$_3$ metric is
\begin{equation}\label{eq:SdS3metric}
 ds^2=-f(r)\,dt_{\rm s}^2+\frac{dr^2}{f(r)}+r^2d\psi_{\rm s}^2,
 \qquad
 f(r)=r_c^2-\frac{r^2}{\ell^2},
 \qquad r_c\in(0,1],
 \qquad \psi_{\rm s}\sim\psi_{\rm s}+2\pi.
\end{equation}
Its cosmological horizon and inverse temperature are
\begin{equation}\label{eq:horizon-temperature}
 r_h=\ell r_c,
 \qquad
 \beta=\frac{2\pi\ell}{r_c}.
\end{equation}
For $r_c<1$, the origin is a conical defect.  The coordinate change
\begin{equation}\label{eq:rescale}
 \widehat t=r_c t_{\rm s},
 \qquad
 \widehat r=\frac r{r_c},
 \qquad
 \widehat\psi=r_c\psi_{\rm s}
\end{equation}
puts the metric in the static dS$_3$ form
\begin{equation}\label{eq:dS3static}
 ds^2=-\left(1-\frac{\widehat r^2}{\ell^2}\right)d\widehat t^2
 +\frac{d\widehat r^2}{1-\widehat r^2/\ell^2}
 +\widehat r^2d\widehat\psi^2,
\end{equation}
with $\widehat\psi\sim\widehat\psi+2\pi r_c$.

We specialize to
\begin{equation}\label{eq:integer-orbifold}
 r_c=\frac1q,
 \qquad q\in\mathbb Z_{\geq2},
\end{equation}
so that the spacetime is the finite quotient
$\mathrm{dS}_3/\mathbb Z_q$.  The quotient acts by an angular rotation at
fixed Lorentzian time\footnote{The quotient used here is the untwisted angular quotient at fixed Lorentzian time. Its Euclidean continuation
is correspondingly the singular orbifold with the same fixed conical locus. This should be distinguished from smooth Lens-space quotients of $S^3$, which involve a free action mixing the Euclidean time and angular circles and prepare different static-patch states \cite{Castro:2011xb}.}. This construction is analogous to the quotient
description of BTZ \cite{Banados:1992wn}; point-particle geometries in
three-dimensional de Sitter gravity were studied in
Ref.~\cite{DESER1984405}. Note that a cone of general opening angle
is not a quotient of dS$_3$ by a finite subgroup, and the method of images
used below would not apply to it.

Near $r=0$, setting $R_\perp:=r/r_c$ gives
\begin{equation}\label{eq:defect-local-metric}
 ds^2_{\rm spatial}
 =\left[1+O(R_\perp^2/\ell^2)\right]dR_\perp^2
 +\frac{R_\perp^2}{q^2}d\psi_{\rm s}^2.
\end{equation}
The deficit angle is
\begin{equation}\label{eq:deficit-angle}
 \delta_{\rm def}=2\pi\left(1-\frac1q\right).
\end{equation}
For the calculations below it is convenient to use the global covering-space
coordinates $(t,\theta,\psi)$, with $t\in\mathbb{R}$, $0\le\theta\le\pi$ and
$\psi\sim\psi+2\pi$, in which
\begin{equation}\label{eq:global}
  ds^{2}=\ell^{2}\big[-dt^{2}+\cosh^{2}t\,(d\theta^{2}+\sin^{2}\theta\,d\psi^{2})\big].
\end{equation}
The static and global radial coordinates are related by
$\hat r=\ell\cosh t\sin\theta$, so the cosmological horizon $\hat r=\ell$ sits
at $\cosh t\sin\theta=1$. In these coordinates the quotient acts as
$\psi\sim\psi+2\pi/q$ and its fixed worldlines are the poles $\theta=0,\pi$.

\subsection{Differentiate-dictionary one-point function}\label{sec:summary}
We consider a heavy scalar $\varphi$, of mass $M$, and a light
scalar $\chi$ on a fixed SdS$_3$ background, with interaction
\begin{equation}\label{eq:cubic-interaction}
 S_{\rm int}
 =g\int d^3x\,\sqrt{-g}\,\varphi\chi^2.
\end{equation}
For the heavy principal-series field we write
\begin{equation}\label{eq:heavy-weights}
 \nu:=\sqrt{M^2\ell^2-1},
 \qquad
 \Delta_\pm:=1\pm i\nu,
 \qquad
 \nu\gg1.
\end{equation}

At a finite late-time surface $\Sigma_c$, let $\varphi_c$ and $\chi_c$
denote the boundary values of the two fields. Their Bunch--Davies
wavefunctional is
\begin{equation}\label{eq:two-field-wavefunctional}
 \Psi_c^{\rm BD}[\varphi_c,\chi_c]
 :=\langle\varphi_c,\chi_c;\Sigma_c|\mathrm{BD}\rangle
 =\int_{\substack{\varphi|_{\Sigma_c}=\varphi_c\\
                   \chi|_{\Sigma_c}=\chi_c}}^{\rm BD}
 \mathcal D\varphi\,\mathcal D\chi\;e^{iS}.
\end{equation}
After adding local counterterms and removing the cutoff, we denote the
renormalized heavy field source by $\mathcal{J}$ and the light field source by
$\chi_{(0)}$. For a chosen heavy field bulk--boundary kernel $K(P;x)$,
let $\varphi_{c,K}[\mathcal{J}]$ be the regulated heavy field datum corresponding to
$\mathcal{J}$. The renormalized wavefunction coefficient and the associated
differentiate-dictionary one-point function are
\begin{equation}\label{eq:differentiate-definition}
 \mathcal W_K[\mathcal{J},\chi_{(0)}]
 :=\operatorname*{Ren}_{\Sigma_c\to\mathcal I^+}
 \left[-i\log
 \Psi_c^{\rm BD}\!\left[
 \varphi_{c,K}[\mathcal{J}],\chi_c[\chi_{(0)}]\right]\right],
 \qquad
 \langle\mathcal O(P)\rangle_K
 :=\left.
 \frac{\delta\mathcal W_K}{\delta \mathcal{J}(P)}
 \right|_{\mathcal{J}=\chi_{(0)}=0},
\end{equation}
where $\operatorname{Ren}$ denotes the cutoff removal after the addition of
local counterterms, and $\chi_c[\chi_{(0)}]$ is the regulated datum
corresponding to the renormalized light field source. The subscript records the kernel used to build the regulated datum from 
$\mathcal
J$; different source conventions give different 
$\mathcal W$, as we make precise in Section II C. Thus the
``one-point function'' considered here is the boundary observable defined
by differentiating the Bunch--Davies wavefunctional; it is not a bulk
in--in expectation value.

At first order in $g$, the perturbative expansion contains $\chi^2(x)$ at
the vertex $g\varphi\chi^2$.  Performing the Gaussian integral over
$\chi$ replaces this pair by its two-point function at coincidence.
The light field factor entering the cubic vertex is the ordinary
renormalized coincidence limit
\begin{equation}\label{eq:chi2-ren-summary}
 \langle\chi^2(x)\rangle_{\rm ren}
 :=\lim_{x'\to x}\left[
 G_D^{(q)}(x,x')-G_{\rm Had}(x,x')\right],
\end{equation}
Here, $G_D^{(q)}(x,x')$ denotes the late-time light field Green function on
SdS$_3$ while $G_{\rm Had}$ is the local Hadamard parametrix containing the
universal short-distance singularity.

$G_D^{(q)}(x,x')$ is obtained by first imposing a homogeneous Dirichlet
condition at the finite final surface $\Sigma_c$ and then taking
$\Sigma_c\to\mathcal I^+$ at fixed interior points. The superscript $(q)$
labels the quotient. The subscript $D$ records this finite-cutoff Dirichlet prescription.

At first order in $g$, the corresponding contribution to the wavefunction
coefficient is then:
\begin{equation}\label{eq:wavefunctional-first-order}
 \mathcal W_K^{(1)}[\mathcal{J},0]
 =g\int d^3x\,\sqrt{-g(x)}\;
 \varphi_K[\mathcal{J}](x)\,
 \langle\chi^2(x)\rangle_{\rm ren},
 \qquad
 \varphi_K[\mathcal{J}](x)
 =\int_{S^2/\mathbb Z_q}d\Omega_{P'}\,K(P';x)\mathcal{J}(P').
\end{equation}
Here $P'$ is a boundary point and $d\Omega_{P'}$ is the unit-sphere
measure. Here the integral is over the quotient space $S^2/\mathbb Z_q$. The kernel $K(P';x)$ is the quotient kernel (constructed below in \eqref{eq:heavy-kernel-image-sum}).

Differentiating with respect
to $\mathcal{J}$ gives
\begin{equation}\label{eq:holoformula}
 \langle\mathcal O(P)\rangle_K
 =g\int d^3x\,\sqrt{-g(x)}\;
 K(P;x)\langle\chi^2(x)\rangle_{\rm ren}.
\end{equation}
\subsection{Choice of kernel}\label{sec:kernels}
We now specify the three source conventions used in the paper. 
With bulk point at
$x=(t,\theta,\psi)$ and the boundary point at $(\theta_P,\psi_P=0)$, the de Sitter
bulk--boundary invariant is
\begin{equation}\label{eq:u-summary}
 u(P;x)
 :=2\!\left[\sinh t-\cosh t\left(
 \cos\theta\cos\theta_P+
 \sin\theta\sin\theta_P\cos\psi\right)\right].
\end{equation}
Here we have used the global coordinates \eqref{eq:global}.
The zero locus of $u(P;x)$ is the bulk light cone of the boundary insertion. On the
single-cover hyperbolic section $t=\rho+i\pi/2$, $\rho\geq0$, write
$x_E=(\rho,\theta,\psi)$. The corresponding positive EAdS invariant is
\begin{equation}\label{eq:U-summary}
 U(P;x_E)
 :=2\!\left[\cosh\rho-\sinh\rho\left(
 \cos\theta\cos\theta_P+
 \sin\theta\sin\theta_P\cos\psi\right)\right].
\end{equation}
Thus $u$ is adapted to Lorentzian boundary values, while $U$ is adapted to
the regular Euclidean continuation. We suppress their arguments when no
ambiguity can arise. With the standard EAdS$_3$ unit-source normalization
\cite{Freedman:1998tz},
\begin{equation}\label{eq:eads-kernel-normalization}
 K_E^{(\Delta)}(P;x_E)=c_\Delta U(P;x_E)^{-\Delta},
 \qquad
 c_\Delta:=\frac{\Gamma(\Delta)}{\pi\Gamma(\Delta-1)}
 =\frac{\Delta-1}{\pi}.
\end{equation}
This formula is initially defined for $\re\Delta>1$ and then analytically
continued to the principal series. In particular,
\begin{equation}\label{eq:principal-series-kernel-coefficients}
 c_{\Delta_+}=\frac{i\nu}{\pi},
 \qquad
 c_{\Delta_-}=-\frac{i\nu}{\pi}.
\end{equation}

We first define the upper Euclidean source $\mathcal J_{E,+}$, the
coefficient of the $\Delta_+$ mode normalized as in
\eqref{eq:eads-kernel-normalization}, where $U>0$ makes $U^{-\Delta_+}$
single-valued. The coefficients $\mathcal J_+$ and $\mathcal J_-$ are the
coefficients of the same falloffs normalized instead on the real
Lorentzian section, where $u$ changes sign across the light cone and the
branch is fixed by the $+i0$ and $-i0$ boundary values respectively. The
generic source $\mathcal J$ in \eqref{eq:differentiate-definition} means
the coefficient appropriate to the kernel being used.

The Hartle--Hawking kernel \cite{Hartle:1983ai} is obtained by continuing
the Euclidean unit-source kernel through the regular upper cap:
\begin{equation}\label{eq:HH-kernel-summary}
 \left.K_{\rm HH}(P;x)\right|_{t=\rho+i\pi/2}
 :=c_{\Delta_+}U(P;x_E)^{-\Delta_+}.
\end{equation}
It is the kernel used when differentiating with respect to
$\mathcal J_{E,+}$.
We separately define the two Lorentzian unit-source bases by
\begin{equation}\label{eq:lorentzian-kernels}
 K_L(P;x):=c_{\Delta_+}(u(P;x)+i0)^{-\Delta_+},
 \qquad
 K_L^*(P;x):=c_{\Delta_-}(u(P;x)-i0)^{-\Delta_-}.
\end{equation}
Here $K_L$ is paired with $\mathcal J_+$ and the upper preparation.
The kernel $K_L^*$ is paired with $\mathcal J_-$ and the conjugate lower
preparation. The displayed $i0$ specifies the distributional boundary
value across the Lorentzian light cone $u=0$.

The three definitions use the same coefficients $c_{\Delta_\pm}$, but
``unit source'' refers to three different asymptotic bases. Their relation
is fixed by analytic continuation. Since
$u(\rho+i\pi/2)=iU(\rho)$, continuation from the upper hyperbolic section
gives
\begin{equation}\label{eq:kernel-hh-lorentzian}
 K_{\rm HH}=e^{i\pi\Delta_+/2}K_L
 =i e^{-\pi\nu/2}K_L.
\end{equation}
Define
\begin{equation}\label{eq:upper-kernel-connection-coefficient}
 f_+:=e^{i\pi\Delta_+/2}=i e^{-\pi\nu/2}.
\end{equation}
The same  bulk field configuration may then be written in either basis as
\begin{equation}\label{eq:upper-source-conversion}
 K_{\rm HH}\mathcal J_{E,+}=K_L\mathcal J_+,
 \qquad
 \mathcal J_+=f_+\mathcal J_{E,+}.
\end{equation}
At linear order in the heavy source, the chain rule gives
\begin{equation}\label{eq:upper-onepoint-conversion}
 \langle\mathcal O(P)\rangle_{K_{\rm HH}}
 =f_+\langle\mathcal O(P)\rangle_{K_L}.
\end{equation}
For the conjugate relation, let $\mathcal J_{E,-}$ denote the coefficient
of the Euclidean-normalized $\Delta_-$ mode continued through the lower cap,
$t=\rho-i\pi/2$. Since $u=-iU$ there, define
\begin{equation}\label{eq:lower-kernel-connection-coefficient}
 f_-:=e^{-i\pi\Delta_-/2}=\overline{f_+}
 =-i e^{-\pi\nu/2}.
\end{equation}
The lower-cap and Lorentzian source coefficients for the same  bulk field configuration
then obey
\begin{equation}\label{eq:lower-source-conversion}
 c_{\Delta_-}U^{-\Delta_-}\mathcal J_{E,-}
 =K_L^*\mathcal J_-,
 \qquad
 \mathcal J_-=f_-\mathcal J_{E,-}.
\end{equation}
If $\langle\mathcal O\rangle_{E,-}$ denotes differentiation with respect
to $\mathcal J_{E,-}$, the corresponding chain-rule relation is
\begin{equation}\label{eq:lower-onepoint-conversion}
 \langle\mathcal O(P)\rangle_{E,-}
 =f_-\langle\mathcal O(P)\rangle^{\rm lower}_{K_L^*}.
\end{equation}
Thus $\mathcal J_{E,-}$ labels a distinct lower-cap Euclidean basis. It is
not notation for the numerical complex conjugate of $\mathcal J_{E,+}$.
The two become conjugate only when the boundary data themselves obey the
corresponding reality condition.

Equations~\eqref{eq:upper-source-conversion} and
\eqref{eq:lower-source-conversion} are changes of source basis. Identifying
the Euclidean and Lorentzian source coefficients numerically while also
retaining $f_\pm$ would compare different  bulk field configurations. The factors change
the angle-independent weights assigned to the one-point coefficients. They
do not change the bulk saddles or their $\theta_P$ dependence.

This is also where the de Sitter normalization question differs from the
usual AdS one. In AdS/CFT, the independently defined boundary CFT fixes the
operator and source convention. Here the differentiate dictionary does not
supply an additional canonical normalization. A
dual description would be needed to make one of the bases canonical. We therefore state
the source basis whenever comparing the three kernels.

\subsection{The renormalized light-field two-point function}

We now turn to the computation of $\langle\chi^2(x)\rangle_{\rm ren}$. We start from the covering-space and
quotient Green functions  $G_D^{(1)}(x,x')$ and $G_D^{(q)}(x,x')$ introduced in \ref{sec:summary}. As mentioned there, the initial condition is the
Bunch--Davies one; $G_D^{(q)}$ is therefore not the unrestricted
Bunch--Davies Wightman function.

On the covering dS$_3$ geometry the global angle has period $2\pi$, and
the generator of the quotient acts as
\begin{equation}\label{eq:quotient-generator}
 \gamma:(t,\theta,\psi)\longmapsto
 \left(t,\theta,\psi+\frac{2\pi}{q}\right).
\end{equation}
For an ordinary quotient-invariant scalar, the method of images gives
\begin{equation}\label{eq:dirichlet-quotient-image-sum}
 G_D^{(q)}(x,x')
 =\sum_{n=0}^{q-1}G_D^{(1)}(x,\gamma^n x').
\end{equation}
The ordinary point-split renormalized coincidence limit is
\begin{equation}\label{eq:chi2-ren-main}
 \left\langle\chi^2(x)\right\rangle_{\rm ren}
 :=\lim_{x'\to x}\left[G_D^{(q)}(x,x')-G_{\rm Had}(x,x')\right],
\end{equation}
where $G_{\rm Had}$ is the local Hadamard parametrix.  Hence
\begin{align}
 \left\langle\chi^2(x)\right\rangle_{\rm ren}
 &=\left\langle\chi^2(x)\right\rangle_{\rm ren}^{(1)}
   +\sum_{n=1}^{q-1}G_D^{(1)}(x,\gamma^n x),
 \label{eq:renormalized-image-contraction}\\
 \left\langle\chi^2(x)\right\rangle_{\rm ren}^{(1)}
 &:=\lim_{x'\to x}\left[G_D^{(1)}(x,x')-G_{\rm Had}(x,x')\right].
 \label{eq:cover-renormalized-coincidence}
\end{align}
Thus $\langle\chi^2\rangle_{\rm ren}$ is the standard renormalized
coincident two-point function on the quotient.  The image sum is simply a
convenient point-split evaluation of it.  This coincidence limit is first
taken away from the quotient fixed worldlines, where every nonidentity
image is separated; the near-defect limit $v\to0$ is taken afterward.
Rotational symmetry makes the result independent of $\psi$.

We choose
\begin{equation}\label{eq:critical-light-mass}
 m_\chi^2\ell^2=1.
\end{equation}
This is the critical value $\nu_\chi=0$, at which the two weights
$1\pm i\nu_\chi$ coincide. Its Euclidean continuation is the
Breitenlohner--Freedman value \eqref{eq:critical-bf-value} in EAdS$_3$. The
covering-space Green function is elementary at any mass, being the
continuation of $e^{-(\Delta_\chi-1)\sigma}/(4\pi\ell_E\sinh\sigma)$, with
$\sigma$ the geodesic separation. The critical value is special in that the
exponential factor disappears, leaving an algebraic function of $\sigma$.
As a result, the image sum
\eqref{eq:critical-exact-quotient-profile} is a finite sum of square roots.
The branch loci of the vertex integrand form a finite set that can be
enumerated and shown not to obstruct the contour deformation
(Appendix~\ref{app2}). Further, the near-defect expansion
\eqref{eq:critical-exact-near-tip} has no $O(v^0)$ term, so corrections to
the heavy-mass limit proceed in powers of $\nu^{-2}$.  

The degenerate weight also carries a cost. Because the indicial roots
coincide, a logarithmic solution appears and the Green function is not
fixed by its coincidence singularity alone. Imposing the final Dirichlet
condition at finite cutoff and then removing the cutoff suppresses the
logarithmic boundary solution and selects the no-logarithm Green function.
Appendix~\ref{app1} derives this by an explicit mode construction. Here we
quote the result.

For dS$_3$ embedding coordinates $X^A$, $A=0,1,2,3$, define
\begin{equation}\label{eq:ds-invariant-s}
 s(x,x'):=\frac{\eta_{AB}X^A(x)X^B(x')}{\ell^2},
 \qquad
 \eta_{AB}=\operatorname{diag}(-,+,+,+),
 \qquad
 s(x,x)=1.
\end{equation}
The covering-space Green function is
\begin{equation}\label{eq:critical-dirichlet-invariant}
 G_D^{(1)}(s)=\frac{1}{4\pi\ell\sqrt{1-s^2}}.
\end{equation}
The square root is positive for real $-1<s<1$. Elsewhere it is defined by
continuation from that domain along the upper edge $\operatorname{Im}t>0$,
on which the Euclidean cap lies; we refer to this throughout as the
Bunch--Davies prescription, and \eqref{eq:critical-profile-BD-branch}
records it explicitly for the image combination. It is the same boundary
value that supplies the $+i0$ in \eqref{eq:lorentzian-kernels}.  At coincidence,
$G_D^{(1)}=1/(4\pi d)+O(d)$, where $d$ is the local proper distance.
Consequently the covering-space finite part in
\eqref{eq:cover-renormalized-coincidence} vanishes in the minimal
Hadamard scheme used here.

For the $n$th quotient image, set
\begin{equation}\label{eq:critical-image-data}
 a_n:=\sin\frac{\pi n}{q},
 \qquad
 v:=\frac{\widehat r}{\ell}=\cosh t\,\sin\theta,
 \qquad
 s_n(x):=s(x,\gamma^n x)=1-2v^2a_n^2.
\end{equation}
Equations~\eqref{eq:renormalized-image-contraction} and
\eqref{eq:critical-dirichlet-invariant} therefore give
\begin{equation}\label{eq:critical-exact-quotient-profile}
 \left\langle\chi^2(v)\right\rangle_{\rm ren}
 =\frac{1}{8\pi\ell v}
 \sum_{n=1}^{q-1}
 \frac{1}{a_n\sqrt{1-v^2a_n^2}}.
\end{equation}
For sufficiently small positive $v$, all roots in
\eqref{eq:critical-exact-quotient-profile} are positive.  Elsewhere the
expression is defined by continuation from that domain with the same
Bunch--Davies prescription.

Defining
\begin{equation}\label{eq:Cq-def}
 C_q:=\frac{1}{8\pi}\sum_{n=1}^{q-1}\csc\frac{\pi n}{q},
\end{equation}
the near-defect series is
\begin{align}
 \left\langle\chi^2(v)\right\rangle_{\rm ren}
 &=\frac{C_q}{\ell v}
 +\frac{v}{16\pi\ell}\cot\frac{\pi}{2q}
 \notag\\
 &\quad
 +\frac{3v^3}{256\pi\ell}
 \left(3\cot\frac{\pi}{2q}-\cot\frac{3\pi}{2q}\right)
 +O(v^5/\ell).
 \label{eq:critical-exact-near-tip}
\end{align}
In particular, the regular part begins at order $v$, rather than at order
$v^0$.  Appendix~\ref{app1} gives the series derivation.

Finally, on the upper hyperbolic section and its conjugate,
$t=\rho\pm i\pi/2$,
\begin{equation}\label{eq:critical-local-analyticity}
 1-v^2a_n^2
 =1+a_n^2\sinh^2\rho\,\sin^2\theta>0
\end{equation}
for real $\rho$ and $\theta$. Hence the explicit $1/v$ pole is the only
light-field singularity in the near-axis region controlling the
large-mass expansion.

\section{The Bunch--Davies contour and the heavy-mass saddle}\label{op}

We now evaluate the one-point function using the exact light-field profile
in \eqref{eq:critical-exact-quotient-profile}. We first specify the
spacetime integration associated with the Bunch--Davies wavefunctional. The
resulting integral is then reduced to a single radial transform, whose
heavy-mass limit identifies the saddle selected by the prepared state.

At first order in $g$, the interaction vertex is integrated over the same
history that defines the Bunch--Davies wavefunctional. A regular Euclidean
cap prepares the state. Lorentzian evolution carries it to the final
surface at $t=t_c$. In global time this history is
\begin{equation}\label{eq:BD-preparation-cycle}
 \mathcal C_{\rm BD}:
 \qquad
 t=\frac{i\pi}{2}\longrightarrow0\longrightarrow t_c.
\end{equation}
We refer to $\mathcal C_{\rm BD}$ as the Bunch--Davies contour. Its first
segment is the regular Euclidean cap that prepares the state; its second is
the Lorentzian part of the wavefunction evolution, along which the $+i0$
boundary values are inherited. After the local counterterms at $t_c$ have
been included, the limit $t_c\to\infty$ is taken \cite{Maldacena:2002vr,
Anninos:2014lwa}. A past-$i\epsilon$ representation is equivalent when it
implements the same interacting Bunch--Davies state preparation.

The $n$th quotient image shifts the azimuth by
$\delta_n=2\pi n/q$. Its heavy-field invariant is
\begin{equation}\label{eq:image-invariant}
u_n:=2\!\left[\sinh t-\cosh t\left(
 \cos\theta\cos\theta_P+
 \sin\theta\sin\theta_P\cos(\psi+\delta_n)\right)\right].
\end{equation}
Thus the quotient kernel is the finite image sum
\begin{equation}\label{eq:heavy-kernel-image-sum}
 K_L^{(q)}(P;x)
 =c_{\Delta_+}\sum_{n=0}^{q-1}(u_n+i0)^{-\Delta_+}.
\end{equation}
The light-field profile is invariant under the quotient rotation. Each
term in \eqref{eq:heavy-kernel-image-sum} therefore maps the fundamental
azimuthal interval to an adjacent interval. Unfolding the image sum gives
\begin{align}
 \left\langle\mathcal O(\theta_P)\right\rangle_{K_L}
 &=g\ell^3
 \int_{\mathcal C_{\rm BD}}dt\int_0^\pi d\theta
 \int_0^{2\pi/q}d\psi\;
 \cosh^2t\,\sin\theta\,
 K_L^{(q)}(P;x)
 \left\langle\chi^2(t,\theta)\right\rangle_{\rm ren}
 \notag\\
 &=g\ell^3c_{\Delta_+}
 \int_{\mathcal C_{\rm BD}}dt\int_0^\pi d\theta
 \int_0^{2\pi}d\psi\;
 \cosh^2t\,\sin\theta\,
 (u+i0)^{-\Delta_+}
 \left\langle\chi^2(t,\theta)\right\rangle_{\rm ren}.
 \label{eq:onepoint-unfolded}
\end{align}
Here $u:=u_0$. The $+i0$ is inherited from the Lorentzian part of
$\mathcal C_{\rm BD}$. After unfolding, all quotient dependence is carried
by the exact light-field profile.

\subsection{Contour deformation and exact radial reduction}

The integral \eqref{eq:onepoint-unfolded} runs over the two-segment contour
$\mathcal C_{\rm BD}$ and over the two angles. We evaluate it in two steps.
The contour is first moved onto a single hyperbolic section, on which the
geometry is $H^3$ and the integrand is real-analytic. The angular
integrations are then performed in coordinates adapted to the quotient
axis, which is possible because the light-field profile
\eqref{eq:critical-exact-quotient-profile} depends only on the distance
from that axis. A single radial integral survives.

For the first step, let $\mathcal F(t)$ denote the angularly integrated
density in \eqref{eq:onepoint-unfolded}. Appendix~\ref{app2} shows that
neither a zero of the heavy invariant nor a reflected light-field branch
point lies in the open strip $0<\operatorname{Im}t<\pi/2$, and that the
connecting edge at large $\operatorname{Re}t$ vanishes. Hence
\begin{equation}\label{eq:BD-H3-deformation}
 \int_{\mathcal C_{\rm BD}}dt\,\mathcal F(t)
 =\int_0^\infty d\rho\,
 \mathcal F\!\left(\rho+\frac{i\pi}{2}\right),
\end{equation}
with orientation $\rho:0\to\infty$. The singularities that remain on the
Lorentzian boundary of the strip---the light cone of the insertion and the
reflected thresholds \eqref{eq:critical-profile-BD-branch}---are locally
integrable and do not pinch the deformation. The identity is established
in the regulating domain $0<\operatorname{Re}\Delta<1$ and continued to
$\Delta=\Delta_+$.

On the section $t=\rho+i\pi/2$ the invariants continue to
\begin{equation}\label{eq:upper-H3-data}
 u=iU,
 \qquad
 v=i\sinh\rho\,\sin\theta,
 \qquad
 1-a_n^2v^2=1+a_n^2\sinh^2\rho\,\sin^2\theta>0,
\end{equation}
so that continuation from the Euclidean cap makes $U$ positive and every
light-field square root positive, and supplies an overall factor
$ie^{-i\pi\Delta_+/2}$. It is convenient to work with
\begin{equation}\label{eq:varrho-definition}
 \varrho:=-iv=\sinh\rho\,\sin\theta,
\end{equation}
the transverse geodesic radius measured from the quotient axis, so that
$\varrho=0$ is the axis itself.

For the second step we use coordinates $(y,\varrho,\psi)$ on $H^3$ in which
$y$ runs along the quotient axis and $\varrho$ is the transverse radius,
the two fixed worldlines $\theta=0$ and $\theta=\pi$ becoming the halves
$y>0$ and $y<0$ of one complete axis. These coordinates are adapted to both
factors of the integrand: the light-field profile is independent of $y$ and
$\psi$, while $U$ factorizes so that, after a shift of $y$, the insertion
angle appears only through an overall power of $\sin\theta_P$. Both
integrations can therefore be carried out exactly, and they produce a
Legendre function of $\varrho$. The result, derived in
Appendix~\ref{app2}, is
\begin{align}
 \left\langle\mathcal O(\theta_P)\right\rangle_{K_L}
 &=i e^{-i\pi\Delta_+/2}
 \frac{g\ell^2c_{\Delta_+}}8
 \frac{\Gamma(\Delta_+/2)^2}{\Gamma(\Delta_+)}
 (\sin\theta_P)^{-\Delta_+}
 \sum_{n=1}^{q-1}\frac{\mathcal R_{\Delta_+}(a_n)}{a_n},
 \label{eq:exact-BD-onepoint}\\
 \mathcal R_\Delta(a)
 &:=\int_0^\infty d\varrho\,
 \frac{\mathrm P_{\Delta/2-1}(1+2\varrho^2)}
 {\sqrt{1+a^2\varrho^2}},
 \qquad
 a_n=\sin\frac{\pi n}{q}.
 \label{eq:RDelta-definition}
\end{align}

Here $\mathrm P_\lambda$ is the Legendre function of the first kind, and
$\mathcal R_\Delta(a_n)$ is the remaining transverse integral for the $n$th
light-field image; it retains the full radial profile, no approximation
having been made. Its integrand decays as $\varrho^{-2}$, so the integral
converges absolutely on the principal series. Because the heavy image sum
has already unfolded the quotient angular domain, and because the two fixed
worldlines have become the two halves of a single axis, no residual factor
of $q$ or factor of two appears. All dependence on the insertion angle has
factorized into $(\sin\theta_P)^{-\Delta_+}$, and the expression is regular
at $\theta_P=\pi/2$; insertions at the fixed endpoints $\theta_P=0,\pi$
remain excluded.

One assumption enters here. The Green function used above is the
fixed-interior limit \eqref{eq:late-time-dirichlet-limit}, and moving that
limit through the vertex integral is not justified by pointwise convergence
alone, since the finite-cutoff Dirichlet correction is nonuniform in a thin
layer next to the final surface. Appendix~\ref{app2} gives power counting
that controls the fixed-mode part of this layer; the identification of
\eqref{eq:exact-BD-onepoint} with the renormalized late-time coefficient
assumes, as stated there, that the remaining high-angular-momentum boundary
layer is local and is removed by the standard wavefunctional counterterms.
%%%%%%%

\subsection{Heavy-mass limit and saddle selection}

Everything now depends on the large-$\nu$ behaviour of
$\mathcal R_{1+i\nu}(a)$. Appendix~\ref{app2} rewrites the transform as an
exact cosine integral and derives, for fixed $a>0$,
\begin{equation}\label{eq:RDelta-leading-asymptotic}
 \mathcal R_{1+i\nu}(a)
 =\frac1\nu\left[1+\frac{a^2}{2\nu^2}
 +O(\nu^{-4})\right].
\end{equation}

That this is a power of $\nu$ rather than something smaller is the
substance of the result. A rapidly oscillating integrand with no interior
stationary point would contribute beyond all orders; the $1/\nu$ comes
entirely from the endpoint $\varrho=0$, over a region shrinking as
$\varrho=O(\nu^{-1})$, as the scaling limit \eqref{eq:conical-Bessel-limit}
at fixed $\nu\varrho$ makes explicit. That endpoint is the quotient axis,
and what allows it to contribute is the near-defect pole. The factor
$\varrho$ in the transverse measure would by itself suppress the endpoint;
the $1/v$ singularity of \eqref{eq:critical-exact-near-tip} cancels it
exactly, leaving the finite measure \eqref{eq:H3-reduced-measure}. The
regular part of the source begins at order $v$ and is therefore suppressed
by two further powers of $\nu$. The defect thus controls the leading
heavy-mass behaviour---not as an estimate weighed against competing
regions, but as a consequence of \eqref{eq:RDelta-leading-asymptotic},
whose derivation accounts for the whole of the convergent integral
\eqref{eq:RDelta-definition}.

Assembling \eqref{eq:exact-BD-onepoint} requires one further large-$\nu$
input,
\begin{equation}
 \frac{\Gamma((1+i\nu)/2)^2}{\Gamma(1+i\nu)}
 =\sqrt{\frac{2\pi}{\nu}}\,
 e^{-i\pi/4}e^{-i\nu\log 2}
 \left[1+O(\nu^{-1})\right],
 \label{eq:heavy-gamma-ratio}
\end{equation}
whose phase $e^{-i\nu\log 2}$ is what completes $\log(2\sin\theta_P)$ in
the exponent. Collecting the numerical factors produced by
\eqref{eq:RDelta-leading-asymptotic}, \eqref{eq:heavy-gamma-ratio} and the kernel renormalization into
\begin{equation}\label{eq:Cq-tilde-definition}
 \widetilde C_q:=\sqrt{2\pi}\,e^{i\pi/4}C_q,
\end{equation}
and using $c_{\Delta_+}=i\nu/\pi$ together with $M\ell=\nu+O(\nu^{-1})$,
the exact result \eqref{eq:exact-BD-onepoint} becomes
\begin{align}
 \left\langle\mathcal O(\theta_P)\right\rangle_{K_L}
 &\sim
 \frac{\widetilde C_qg\ell^2}
 {\nu^{1/2}\sin\theta_P}
 e^{-i\nu\log(2\sin\theta_P)}e^{\pi\nu/2}
 \left[1+O(\nu^{-1})\right]
 \notag\\
 &\sim
 \frac{\widetilde C_qg\ell^2}
 {\nu^{1/2}\sin\theta_P}e^{-iML_+}.
 \label{eq:KL-heavy-mass-result}
\end{align}

We now locate this contribution on the deformed geometry. Define
\begin{equation}\label{eq:rho0}
 \rho_0:=\log\cot\frac{\theta_P}{2}.
\end{equation}
The factorization \eqref{eq:H3-U-factorization} gives, at $\varrho=0$,
\begin{equation}
 U=2\sin\theta_P\cosh(y-\rho_0).
 \label{eq:axis-U}
\end{equation}
The axial phase is stationary at $y=\rho_0$. The selected configuration
therefore consists of a transverse endpoint at the quotient axis and an
ordinary stationary point along that axis.

For the geodesic comparison, the nearest algebraic critical points in
global coordinates are
\begin{equation}\label{eq:polar-time-saddles}
 t_{{\rm N},\sigma}=\rho_0+\sigma\frac{i\pi}{2},
 \qquad
 t_{{\rm S},\sigma}=-\rho_0+\sigma\frac{i\pi}{2},
 \qquad \sigma=\pm1.
\end{equation}
The single-cover contour \eqref{eq:BD-H3-deformation} contains the upper
representative. For $\rho_0>0$ it lies on the north half of the axis. For
$\rho_0<0$ it lies on the south half. At $\rho_0=0$ it is the integrable
center of the cap. The lower representative belongs to the conjugate
preparation.

For real $g$, let
$\langle\mathcal O\rangle^{\rm lower}_{K_L^*}$ denote the coefficient of
the complex-conjugate wavefunctional, prepared on the lower Euclidean cap.
Complex conjugation then gives
\begin{equation}\label{eq:KL-conjugate-heavy-mass-result}
 \left\langle\mathcal O(\theta_P)\right\rangle^{\rm lower}_{K_L^*}
 \sim
 \frac{\overline{\widetilde C_q}g\ell^2}
 {\nu^{1/2}\sin\theta_P}
 e^{+i\nu\log(2\sin\theta_P)}e^{\pi\nu/2}
 \sim
 \frac{\overline{\widetilde C_q}g\ell^2}
 {\nu^{1/2}\sin\theta_P}e^{+iML_-}.
\end{equation}

Finally, changing from the Lorentzian source $\mathcal J_+$ to the upper
Euclidean source $\mathcal J_{E,+}$ according to
\eqref{eq:upper-onepoint-conversion} gives, with
$\widetilde D_q:=i\widetilde C_q$,
\begin{equation}\label{eq:HH-heavy-mass-result}
 \left\langle\mathcal O(\theta_P)\right\rangle_{K_{\rm HH}}
 \sim
 \frac{\widetilde D_qg\ell^2}
 {\nu^{1/2}\sin\theta_P}
 e^{-i\nu\log(2\sin\theta_P)}.
\end{equation}
The change of source basis removes the angle-independent enhancement of the
$K_L$ coefficient without changing the selected saddle or its angular
dependence.

\section{Complex geodesic action}\label{geodesic}

The heavy-mass exponent in \eqref{eq:KL-heavy-mass-result} contains $L_+$.
We now compare this quantity with the stationary values of a complex
geodesic action. Two logarithms enter the comparison, and their branches
are fixed in different ways. The continued $H^2$ distance fixes the
geometric logarithm, whereas the $+i0$ Bunch--Davies boundary value fixes
the logarithm in the heavy kernel.

Rotational symmetry confines the geodesic to the meridional plane through
the boundary insertion. Away from the polar axis this is a surface of
fixed azimuth $\psi$, with metric
\begin{equation}\label{eq:fixed-azimuth-metric}
 ds^2=\ell^2\left(-dt^2+\cosh^2t\,d\theta^2\right).
\end{equation}
One endpoint approaches $(t,\theta)=(+\infty,\theta_P)$, while the other
lies on the complexified fixed locus of the quotient. Since the position
of the latter endpoint is integrated over in the one-point diagram, its
stationary value must be found by extremizing the geodesic length.

We first continue to the upper hyperbolic section by setting
$t=\rho+i\pi/2$. Equation~\eqref{eq:fixed-azimuth-metric} becomes minus
the metric on $H^2$. Although the usual hyperbolic radius obeys
$\rho\geq0$, the continued north and south worldlines form the two halves
of one complete geodesic axis. It is convenient to parameterize that axis
by a signed coordinate $y$:
\begin{equation}\label{eq:signed-fixed-axis}
 (\rho,\theta)=
 \begin{cases}
  (y,0),&y\geq0,\\
  (-y,\pi),&y\leq0.
 \end{cases}
\end{equation}
Positive and negative values cover the north and south halves,
respectively, and the two meet at $y=0$.

Place the regulated boundary endpoint at
$(\rho_{\rm bdy},\theta_P)$. Its hyperbolic distance from the point $y$ on
the axis satisfies
\begin{equation}\label{eq:H2-distance}
 \cosh\!\left(\frac{d_{H^2}}\ell\right)
 =\cosh y\cosh\rho_{\rm bdy}
 -\sinh y\sinh\rho_{\rm bdy}\cos\theta_P.
\end{equation}
Define
\begin{equation}
 A(y,\theta_P):=\cosh y-\sinh y\cos\theta_P.
 \label{eq:H2-axis-factor}
\end{equation}
For large $\rho_{\rm bdy}$, Eq.~\eqref{eq:H2-distance} gives
\begin{align}
 \cosh\!\left(\frac{d_{H^2}}\ell\right)
 &=\frac{e^{\rho_{\rm bdy}}}{2}A(y,\theta_P)
 +O(e^{-\rho_{\rm bdy}}),
 \notag\\
 d_{H^2}
 &=\ell\rho_{\rm bdy}+\ell\log A(y,\theta_P)
 +O(\ell e^{-2\rho_{\rm bdy}}).
 \label{eq:H2-distance-asymptotic}
\end{align}

The subtraction is defined using the induced angular scale at the cutoff,
which is $\sinh\rho_{\rm bdy}$. The corresponding length counterterm is
\begin{equation}
 \ell\log\sinh\rho_{\rm bdy}
 =\ell(\rho_{\rm bdy}-\log 2)
 +O(\ell e^{-2\rho_{\rm bdy}}).
 \label{eq:H2-length-counterterm}
\end{equation}
This convention fixes the finite additive constant. A constant rescaling
of the boundary conformal metric would shift every regulated length by the
same angle-independent amount. Subtracting
\eqref{eq:H2-length-counterterm} from
\eqref{eq:H2-distance-asymptotic}, we obtain
\begin{align}
 L_E^{\rm reg}(y)
 &:=\lim_{\rho_{\rm bdy}\to\infty}
 \left[d_{H^2}-\ell\log\sinh\rho_{\rm bdy}\right]
 \notag\\
 &=\ell\log\!\left[
 2(\cosh y-\sinh y\cos\theta_P)\right].
 \label{eq:LE-general}
\end{align}

Extremizing the endpoint along the complete axis gives
\begin{equation}
 \partial_yL_E^{\rm reg}=0
 \quad\Longrightarrow\quad
 \tanh y_*=\cos\theta_P.
 \label{eq:H2-axis-extremum}
\end{equation}
The solution is $y_*=\rho_0$, where $\rho_0$ is defined in
\eqref{eq:rho0}. It is unique because $\tanh y$ is monotonic, and it is the
global minimum because $A(y,\theta_P)$ grows at both ends of the axis. At
this point $A(y_*,\theta_P)=\sin\theta_P$, so
\begin{equation}\label{eq:LE-final}
 L_E^{\rm reg}(y_*)=\ell\log(2\sin\theta_P).
\end{equation}
The endpoint lies on the north half for $\theta_P<\pi/2$ and on the south
half for $\theta_P>\pi/2$. At the equator it reaches the common point
$y=0$.

We next return the boundary endpoint to real Lorentzian time. North-pole
coordinates provide a single signed-coordinate representative of the
complexified axis, for which
\begin{equation}\label{eq:north-pole-heavy-invariant}
 u_{\rm N}(t):=2\big(\sinh t-\cosh t\cos\theta_P\big).
\end{equation}
When $y<0$, the same complex point can instead be written in south-pole
coordinates as
$(t=-y+\sigma i\pi/2,\theta=\pi)$.

Place the boundary endpoint at $(T,\theta_P)$, with $T\gg1$. The de Sitter
invariant between this point and $(t,0)$ is
\begin{align}
 s\big((T,\theta_P),(t,0)\big)
 &=-\sinh T\sinh t+\cosh T\cosh t\cos\theta_P
 \notag\\
 &=-\frac{e^T}{4}u_{\rm N}(t)+O(e^{-T}).
 \label{eq:boundary-invariant-asymptotic}
\end{align}

The equation $\cos[D(T,t)/\ell]=s$ does not by itself fix a branch of the
complex distance. We fix it before taking the Lorentzian boundary limit.
On the real upper $H^2$ section we require
$D=-i d_{H^2}$, so that the continued worldline action satisfies
$iD=d_{H^2}>0$, and we then continue this branch through the same upper
analytic domain. Denoting the resulting logarithm by
$\log_{\rm geom}$, its large-$s$ behavior is
\begin{equation}
 D(T,t)=-i\ell\log_{\rm geom}(2s)+O(\ell s^{-2}).
 \label{eq:dS-distance-large-s}
\end{equation}
With $\mathcal L(T,t):=iD(T,t)$, Eq.~\eqref{eq:boundary-invariant-asymptotic}
then gives
\begin{equation}
 \mathcal L(T,t)
 =\ell(T-\log 2)+\ell\log_{\rm geom}[-u_{\rm N}(t)]
 +O(\ell e^{-2T}).
 \label{eq:Lorentzian-bare-action}
\end{equation}
The same boundary conformal representative is used on the Lorentzian
side, where the angular scale of the late-time cutoff is $\cosh T$. Since
$\log\cosh T=T-\log 2+O(e^{-2T})$, subtracting this boundary term yields
\begin{equation}\label{eq:boundary-anchored-length}
 L_{\rm geom}^{\rm reg}(t)
 =\ell\log_{\rm geom}[-u_{\rm N}(t)].
\end{equation}

The endpoint equation $u_{\rm N}'(t)=0$ has the two nearest solutions
$t_{{\rm N},\sigma}$ given in \eqref{eq:polar-time-saddles}, and at these
points
\begin{equation}\label{eq:geodesic-endpoint-saddles}
 u_{\rm N}\big(t_{{\rm N},\sigma}\big)
 =2\sigma i\sin\theta_P,
 \qquad \sigma=\pm1.
\end{equation}
The branch fixed by the positive $H^2$ distance therefore gives
\begin{equation}\label{eq:critical-complex-lengths}
 L_{\rm geom}^{\rm reg}\big(t_{{\rm N},\sigma}\big)
 =\ell\log(2\sin\theta_P)-\sigma\frac{i\pi\ell}{2}
 =L_{-\sigma}.
\end{equation}

The logarithm in the heavy kernel has a separate branch prescription. On
the upper Bunch--Davies contour, $u=iU$ with $U>0$, which fixes
$\log_{\rm BD}u=\log U+i\pi/2$. The lower preparation fixes its conjugate
boundary value. At the stationary points the two branch assignments are
\begin{equation}\label{eq:kernel-geodesic-label-pairing}
 \ell\log_{\rm BD}u_{\rm N}\big(t_{{\rm N},\sigma}\big)=L_\sigma,
 \qquad
 \ell\log_{\rm geom}[-u_{\rm N}\big(t_{{\rm N},\sigma}\big)]
 =L_{-\sigma}.
\end{equation}
%%%%%
The two logarithms take arguments of opposite sign, and this is not an
independent choice. Equation~\eqref{eq:boundary-invariant-asymptotic} expresses the de Sitter
invariant as $s=-\tfrac{1}{4}e^{T}u_N+O(e^{-T})$, so the geometric logarithm
inherits $-u_N$ from the continued distance while the kernel logarithm carries
$u_N$ itself. At a fixed coordinate representative the two values therefore
differ by exactly $\ell\log(-1)$, and on the branches fixed above this
difference is
\begin{equation}\label{eq:branch-shift}
  \ell\log_{\rm geom}\!\big[-u_N(t_{N,\sigma})\big]
  -\ell\log_{\rm BD}\!\big[u_N(t_{N,\sigma})\big]
  \;=\;-\sigma\, i\pi\ell\;=\;L_{-\sigma}-L_{\sigma}.
\end{equation}
The shift is the separation of the two members of the pair. This is
why the two prescriptions attach the labels to opposite coordinate saddles:
the relative branch offset is fixed, not free. Both calculations produce the
same pair $\{L_+,L_-\}$; the upper $K_L$ coefficient selects $L_+$ through
$\log_{\rm BD}u$, and the conjugate preparation selects $L_-$. The comparison
is between action values, with the branch relation~\eqref{eq:branch-shift}
determined rather than assumed. Appendix~\ref{app3} recovers the same
geometric pair from the $S^3$ continuation.

The two parts of the complex length have a direct geometric
interpretation. At the regulated boundary point the static radial
coordinate is $\widehat r_P=\ell\cosh T\sin\theta_P$, and the magnitude of
the radial proper time from the horizon to this point is
\begin{equation}
 \int_\ell^{\widehat r_P}
 \frac{d\widehat r}{\sqrt{\widehat r^2/\ell^2-1}}
 =\ell\,\operatorname{arccosh}(\cosh T\sin\theta_P).
 \label{eq:boundary-horizon-bare-time}
\end{equation}
After applying the same boundary subtraction, this becomes
\begin{equation}\label{eq:boundary-horizon-time}
 \lim_{T\to\infty}\left[
 \ell\,\operatorname{arccosh}(\cosh T\sin\theta_P)
 -\ell(T-\log 2)\right]
 =\ell\log(2\sin\theta_P).
\end{equation}
The proper distance from the defect to the cosmological horizon is
\begin{equation}\label{eq:horizon-defect-distance}
 L_{\rm hor-def}
 =\int_0^{\ell r_c}\frac{dr}
 {\sqrt{r_c^2-r^2/\ell^2}}
 =\frac{\pi\ell}{2}.
\end{equation}
Thus $\operatorname{Re}L_\pm$ is the renormalized magnitude of the
boundary-to-horizon radial proper time, whereas
$|\operatorname{Im}L_\pm|$ is the horizon-to-defect distance. In $e^{-iML_\pm}$, the former supplies the phase and the latter sets the
exponential weight. Their roles are interchanged in the AdS  black hole
example.

\section{Discussion}\label{disc}

Grinberg and Maldacena \cite{Grinberg:2020fdj} showed that in AdS/CFT the thermal one-point function of a heavy boundary
operator encodes the length of a complex geodesic running from the
boundary to the black hole singularity. This papers asked whether an analogous
statement can be made in de Sitter space. We found that it can. 

We considered SdS$_3$, where the conical defect plays the role of the singularity and took the one-point function to be defined by the differentiate dictionary at
$\mathcal I^+$. We found that the heavy-mass limit of the one-point function is
governed by a complex saddle whose action is a complex geodesic length
joining future infinity to the defect. The real part is the renormalized
boundary-to-horizon proper time \eqref{eq:boundary-horizon-time} while the
imaginary part is the horizon-to-defect distance
\eqref{eq:horizon-defect-distance}. We showed that this agrees with the independent
geodesic calculation of Section~\ref{geodesic}. The roles of the two segments are thus exchanged from  the AdS case---here the
boundary-to-horizon segment is timelike and supplies the oscillatory
phase while the horizon-to-defect segment is spacelike and sets the
exponential weight. That exchange is expected since 
cosmological and black hole horizons have causally
opposite characters. The physical picture in
both settings is that a source diverging at the singular locus dominates
the heavy-mass limit, so that the probe is effectively localized there and
its action becomes a geodesic length.  

We obtain the result by evaluating the integral \eqref{eq:intro-bulk-integral} exactly. First, the  Bunch--Davies contour
\eqref{eq:BD-preparation-cycle} is deformed onto a single hyperbolic
section. This reduces the integral to the convergent one-dimensional transform
\eqref{eq:RDelta-definition} and its asymptotics follow from an
endpoint expansion (as opposed to a stationary-phase estimate).  

The result depends on the choice of normalization. The coefficient
\eqref{eq:KL-heavy-mass-result} is defined by differentiation with respect
to the Lorentzian source $\mathcal J_+$. Re-expressing the same bulk
profile in the upper Euclidean basis multiplies it by
$f_+$ as in \eqref{eq:upper-onepoint-conversion}, giving
\eqref{eq:HH-heavy-mass-result}. In that case,  only $\operatorname{Re}L_+$
survives. The imaginary part of the geodesic length is therefore visible
in the Lorentzian basis and absorbed into the connection factor
\eqref{eq:upper-kernel-connection-coefficient} in the Euclidean one. The final choice of basis would have to come from the dual theory. 

As noted in the introduction, our computations were simplified by the restriction to finite cyclic quotients.  It is interesting to note how little of the answer depends on the quotient. The
exponent contains no $q$: $\operatorname{Re}L_\pm$ is fixed by the insertion
angle and $|\operatorname{Im}L_\pm|$ by
\eqref{eq:horizon-defect-distance}, which is independent of $r_c$. All
$q$-dependence resides in the prefactor $\widetilde C_q$ and in the
subleading corrections through $a_n=\sin(\pi n/q)$. Based on this, we expect the same exponent should hold for a
general cone angle. The two ingredients that appear in the answer are the coefficient of the
near-defect pole and the
horizon-to-defect distance. The former is fixed by the local Hadamard singularity and the latter by the metric.
Neither require the deficit to be of quotient type. But an
exact treatment would be unavailable, since the finite branch structure that controls the
contour deformation would be replaced by a cut. We leave this for future consideration. 

A further natural extension would be to the case of the general de Sitter black hole. This would be more challenging and the methods of this paper would not go through. We expect that the part about the near-defect saddle would go through. However, the global statement that this
saddle exhausts the leading behaviour need not carry over. We leave this for future consideration. 

The connection between bulk two-point function in the heavy limit and complex geodesic distance in de Sitter has been explored in
\cite{Fischetti:2014uxa,Chapman:2022mqd,Aalsma:2022eru}. It will be interesting to connect these works with ours. 

\begin{acknowledgments}
It is a pleasure to thank Parijat Dey for initial collaboration on the
project and Justin David for discussions. ChatGPT Sol was used to help with the writing of the paper. 
\end{acknowledgments}

\bibliographystyle{unsrt}
\bibliography{main}

\appendix

\section{Critical-mass Dirichlet propagator}\label{app1}
This appendix constructs the light-field Green function used in the main
text and derives its near-defect expansion. Section~\ref{app1:cutoff}
solves the covering-space problem at finite cutoff and shows that removing
the cutoff at fixed interior points selects the no-logarithm solution;
this is the step that distinguishes the wavefunctional Green function from
the unrestricted Bunch--Davies Wightman function, and it is what fixes
\eqref{eq:critical-dirichlet-invariant}. Section~\ref{app1:lorentzian}
gives the Lorentzian form and records the relation
\eqref{eq:critical-dirichlet-wightman-sum} between the two.
Section~\ref{app1:series} performs the quotient image sum, derives the
near-defect series \eqref{eq:critical-exact-near-tip} whose leading $1/v$
term controls the heavy-mass limit, and establishes the large-$v$ falloff
\eqref{eq:critical-profile-large-v} used in the contour estimates of
Appendix~\ref{app2}.

Let $\Sigma_c:=\{t=t_c\}$ be a finite late-time surface.  In the
wavefunctional path integral the final value $\chi_c$ is held fixed, so we
write
\begin{equation}\label{eq:critical-dirichlet-fluctuation}
 \chi=\chi_{\rm cl}[\chi_c]+\delta\chi,
 \qquad
 \delta\chi|_{\Sigma_c}=0.
\end{equation}
The two-point function used at the cubic vertex is therefore the Green
function of $\delta\chi$.  Denote the covering-space
finite-cutoff Green function by $G_{D,\Sigma_c}^{(1)}(x,x')$.  Here the
subscript $D$ records its homogeneous final Dirichlet condition,
$\Sigma_c$ records the cutoff surface, and $(1)$ labels the covering
geometry.  It
satisfies
\begin{equation}\label{eq:wavefunction-dirichlet-green}
 (\Box_x-m_\chi^2)G_{D,\Sigma_c}^{(1)}(x,x')
 =\frac{i\,\delta^{(3)}(x,x')}{\sqrt{-g(x)}},
 \qquad
 G_{D,\Sigma_c}^{(1)}(x,x')\big|_{x\in\Sigma_c}=0,
\end{equation}
with the analogous condition in the second argument and the
Bunch--Davies initial condition.  A general finite-cutoff construction of
such wavefunction Green functions is given in Appendix B of
Ref.~\cite{Anninos:2014lwa}.

We keep $m_\chi^2\ell^2=1$ fixed and define the Green function used in the
main text by
\begin{equation}\label{eq:late-time-dirichlet-limit}
 G_D^{(1)}(x,x')
 :=\lim_{\Sigma_c\to\mathcal I^+}G_{D,\Sigma_c}^{(1)}(x,x'),
\end{equation}
where the limit is taken at fixed interior points.  The near-defect and
heavy-mass limits are taken only afterward.
Appendix~\ref{app2} gives the boundary power counting that permits this
limit to be used in the nonlocal vertex coefficient after the standard
wavefunctional counterterms are included.

\subsection{Finite cutoff and the no-logarithm limit}\label{app1:cutoff}

The Euclidean continuation maps the problem to EAdS$_3$ with positive
radius $\ell_E$ (numerically equal to $\ell$ under the continuation).  The
continued light-field mass $m_{\chi,E}$ is at the
Breitenlohner--Freedman value
\begin{equation}\label{eq:critical-bf-value}
 m_{\chi,E}^2\ell_E^2=-1.
\end{equation}
In global coordinates,
\begin{equation}\label{eq:eads3-global-critical}
 ds_E^2=\ell_E^2\left(d\rho^2+\sinh^2\rho\,d\Omega_2^2\right),
 \qquad 0\leq\rho\leq R_E.
\end{equation}
Here $R_E$ is the Euclidean radial cutoff corresponding to the regulated
final surface.  We denote EAdS points by
$x_E=(\rho,\Omega)$ and $x_E'=(\rho',\Omega')$, with
$\Omega=(\theta,\psi)$.
Expand in normalized spherical harmonics
$Y_{j\mu}$, with
$-\nabla_{S^2}^2Y_{j\mu}=j(j+1)Y_{j\mu}$.  Away from the source, a radial
mode satisfies
\begin{equation}\label{eq:bf-radial-equation}
 f_j''+2\coth\rho\,f_j'
 -j(j+1)\operatorname{csch}^2\!\rho\,f_j+f_j=0.
\end{equation}
Writing $f_j=\operatorname{csch}\rho\,y_j$ reduces this equation to
Legendre's equation.  We denote the no-logarithm and logarithmic basis
modes by
\begin{equation}\label{eq:bf-legendre-basis}
 f_j^{\rm nl}(\rho)
 :=\operatorname{csch}\rho\,\mathrm{P}_j(\coth\rho),
 \qquad
 f_j^{\log}(\rho)
 :=\operatorname{csch}\rho\,\mathrm{Q}_j(\coth\rho).
\end{equation}
The mode $f_j^{\log}=O(\rho^j)$ is regular at the center, while
$f_j^{\rm nl}=O(\rho^{-j-1})$ is singular there.  At large $\rho$,
\begin{align}
 f_j^{\rm nl}(\rho)
 &=2e^{-\rho}\left[1+O(e^{-2\rho})\right],
 \label{eq:bf-no-log-asymptotic}\\
 f_j^{\log}(\rho)
 &=2e^{-\rho}\left[\rho-\mathfrak h_j
 +O(\rho e^{-2\rho})\right],
 \qquad
 \mathfrak h_j:=\sum_{k=1}^j\frac1k,
 \quad \mathfrak h_0:=0.
 \label{eq:bf-log-asymptotic}
\end{align}
Their weighted Wronskian is
\begin{equation}\label{eq:bf-wronskian}
 \sinh^2\rho\left[
 f_j^{\rm nl}(f_j^{\log})'-(f_j^{\rm nl})'f_j^{\log}
 \right]=1.
\end{equation}

The solution that vanishes at the cutoff is
\begin{equation}\label{eq:bf-cutoff-mode}
 f_{j,R_E}^{D}(\rho)
 :=f_j^{\rm nl}(\rho)
 -\frac{f_j^{\rm nl}(R_E)}{f_j^{\log}(R_E)}f_j^{\log}(\rho),
 \qquad f_{j,R_E}^{D}(R_E)=0.
\end{equation}
Let $G_{D,R_E}^E(x_E,x_E')$ denote the Euclidean Dirichlet Green function
on the ball $\rho\leq R_E$, normalized by
$(-\nabla_E^2+m_{\chi,E}^2)G_{D,R_E}^E
=\delta_E^{(3)}(x_E,x_E')/\sqrt{g_E}$, where $\delta_E^{(3)}$ is the
coordinate delta distribution.  With
$\rho_<:=\min(\rho,\rho')$ and $\rho_>:=\max(\rho,\rho')$, it is
\begin{equation}\label{eq:bf-finite-cutoff-green}
 G_{D,R_E}^E(x_E,x_E')=\frac1{\ell_E}
 \sum_{j=0}^{\infty}\sum_{\mu=-j}^{j}
 Y_{j\mu}(\Omega)Y_{j\mu}^*(\Omega')
 f_j^{\log}(\rho_<)f_{j,R_E}^{D}(\rho_>).
\end{equation}
For each fixed $j$,
\begin{equation}\label{eq:bf-cutoff-ratio}
 \frac{f_j^{\rm nl}(R_E)}{f_j^{\log}(R_E)}
 =\frac1{R_E-\mathfrak h_j}
 \left[1+O(R_Ee^{-2R_E})\right]
 \longrightarrow0.
\end{equation}
The fixed-interior limit consequently selects the no-logarithm inverse.
We denote this Euclidean limit by
\begin{equation}\label{eq:bf-invariant-green}
 G_{\rm nl}^E(\mathfrak d_E)
 :=\lim_{R_E\to\infty}G_{D,R_E}^E(x_E,x_E')
 =\frac1{4\pi\ell_E\sinh\mathfrak d_E},
\end{equation}
where the limit is taken at fixed interior points, distributionally after
angular smearing, and $\mathfrak d_E$ is the dimensionless EAdS$_3$
geodesic distance between $x_E$ and $x_E'$.  It has
the canonical $1/(4\pi d_E)$ coincidence singularity, with
$d_E=\ell_E\mathfrak d_E$, and the pure $e^{-\mathfrak d_E}$ boundary
falloff.  The same selection follows directly from
the global dS modes
$\operatorname{sech}t\,\mathrm{P}_j(\tanh t)$ and
$\operatorname{sech}t\,\mathrm{Q}_j(\tanh t)$.  The second mode contains
the late-time logarithm.  In the exterior solution obtained by adding it
to the no-logarithm mode so as to vanish at $t=t_c$, its coefficient is
$O(t_c^{-1})$.

\subsection{Lorentzian form and relation to the Wightman function}\label{app1:lorentzian}

Continuing both the invariant and the Green-function normalization in
\eqref{eq:bf-invariant-green} with the Bunch--Davies prescription gives
$G_D^{(1)}$ in \eqref{eq:critical-dirichlet-invariant}.  Away from its
singularities, any invariant bisolution $F(s)$ at the critical mass obeys
\begin{equation}\label{eq:critical-invariant-equation}
 (1-s^2)F''-3sF'-F=0.
\end{equation}
The two independent solutions are proportional to
$1/\sqrt{1-s^2}$ and
$\arcsin(s)/\sqrt{1-s^2}$.  The latter contains the critical logarithmic
falloff, so the no-logarithm condition and the canonical coincidence
singularity fix $G_D^{(1)}(s)$.

For real $-1<s<1$, the unrestricted Bunch--Davies Wightman function is
\cite{Bunch:1978yq}
\begin{equation}\label{eq:BD-wightman-real}
 W_{\rm BD}^{(1)}(s)=\frac{\pi-\arccos s}
 {4\pi^2\ell\sqrt{1-s^2}}.
\end{equation}
It follows that
\begin{equation}\label{eq:critical-dirichlet-wightman-sum}
 G_D^{(1)}(s)=W_{\rm BD}^{(1)}(s)+W_{\rm BD}^{(1)}(-s).
\end{equation}
The second term is homogeneous and smooth at coincidence, where it equals
$1/(4\pi^2\ell)$, but has the reflected singularity at $s=-1$.  This is
the critical-mass limit of the future-fast Green function discussed in
Ref.~\cite{AnninosNgStrominger:2011}.

\subsection{Finite quotient and near-defect series}\label{app1:series}

For the image data in \eqref{eq:critical-image-data},
\begin{equation}\label{eq:image-invariant-factorization}
 1-s_n(x)^2=4v^2a_n^2(1-v^2a_n^2).
\end{equation}
Substitution in \eqref{eq:renormalized-image-contraction}, together with
the vanishing covering-space finite part, gives
\eqref{eq:critical-exact-quotient-profile}.  Within its disk of convergence,
\begin{equation}\label{eq:critical-profile-binomial}
 \left\langle\chi^2(v)\right\rangle_{\rm ren}
 =\frac1{8\pi\ell v}\sum_{k=0}^{\infty}
 \frac{\binom{2k}{k}}{4^k}v^{2k}
 \sum_{n=1}^{q-1}a_n^{2k-1}.
\end{equation}
The identities
\begin{equation}\label{eq:critical-odd-sine-sums}
 \sum_{n=1}^{q-1}a_n=\cot\frac{\pi}{2q},
 \qquad
 \sum_{n=1}^{q-1}a_n^3
 =\frac14\left(3\cot\frac{\pi}{2q}
 -\cot\frac{3\pi}{2q}\right)
\end{equation}
produce the series \eqref{eq:critical-exact-near-tip}.  The expansion
contains only odd powers after its leading $1/v$ term.

For comparison, set $b_n:=va_n<1$.  The boundary-condition-dependent
homogeneous term has the expansion
\begin{equation}\label{eq:critical-antipodal-expansion}
 W_{\rm BD}^{(1)}\!\left(-s_n(x)\right)
 =\frac1{4\pi^2\ell}
 +\frac{b_n^2}{6\pi^2\ell}+O(b_n^4/\ell).
\end{equation}
Together with $\sum_{n=1}^{q-1}a_n^2=q/2$, this cancels the constant and
$v^2$ terms of the unrestricted Wightman image sum, while leaving the
$1/v$ and $v$ coefficients unchanged.  Each exact image also has a
reflected square-root locus at $v a_n=1$.  Equation
\eqref{eq:critical-local-analyticity} shows directly that these
loci do not enter the local polar saddle neighborhoods.

On the Lorentzian edge inherited from the upper Bunch--Davies preparation,
the square-root boundary value is
\begin{equation}\label{eq:critical-profile-BD-branch}
 \left[1-a_n^2v^2\right]_{\rm BD}^{-1/2}
 =\left[1-a_n^2\cosh^2t\sin^2\theta-i0\,\sinh t\right]^{-1/2}.
\end{equation}
Using
$\sum_{n=1}^{q-1}\csc^2(\pi n/q)=(q^2-1)/3$, the exact profile has the
large-$v$ expansion
\begin{equation}\label{eq:critical-profile-large-v}
 \left\langle\chi^2(v)\right\rangle_{\rm ren}
 =\frac{i(q^2-1)}{24\pi\ell v^2}
 +O(v^{-4}/\ell).
\end{equation}
Here the displayed sign is the future-tail sign; the conjugate preparation
gives its complex conjugate. At fixed angles away from the poles and
light-cone loci, this falloff makes the integrand density in
\eqref{eq:onepoint-unfolded} decay as $O(e^{-t})$.

\section{Exact reduction of the Bunch--Davies integral}\label{app2}
This appendix supplies the four technical inputs used in
Section~\ref{op}. Section~\ref{app2:deform} proves the contour deformation
\eqref{eq:BD-H3-deformation} by showing that no zero of the heavy invariant
and no reflected light-field branch point enters the strip
$0<\operatorname{Im}t<\pi/2$, and that the closing edge
\eqref{eq:future-endcap-u} vanishes as $\operatorname{Re}t\to\infty$.
Section~\ref{app2:cylindrical} introduces coordinates adapted to the
quotient axis, evaluates the axial and azimuthal integrations in
\eqref{eq:H3-orbit-integral}, and thereby reduces the vertex integral to
the one-dimensional transform \eqref{eq:RDelta-definition}, whose absolute
convergence on the principal series is established there.
Section~\ref{app2:asymptotics} derives the large-mass expansion
\eqref{eq:RDelta-leading-asymptotic}, and gives the closed form
\eqref{eq:RDelta-q2} at $q=2$ as a finite-mass check of the whole
reduction. Section~\ref{app2:cutoff} states the boundary power counting
and the counterterm prescription under which the fixed-interior Green
function \eqref{eq:late-time-dirichlet-limit} may be used inside the
vertex integral.
\subsection{Deformation to the upper hyperbolic section}\label{app2:deform}

Write $\Delta$ for a complex heavy weight, initially in the regulating
domain $0<\operatorname{Re}\Delta<1$; the physical value
$\Delta=\Delta_+$ is reached by analytic continuation below. Define the
angular invariant
\begin{equation}\label{eq:angular-invariant-Xi}
 \Xi:=\cos\theta\cos\theta_P+
 \sin\theta\sin\theta_P\cos\psi.
\end{equation}
Then $u=2\cosh t(\tanh t-\Xi)$. For $t=x+iy$ with
$x>0$ and $0<y<\pi/2$,
\begin{equation}\label{eq:imaginary-tanh-strip}
 \operatorname{Im}\tanh(x+iy)
 =\frac{\sin(2y)}{\cosh(2x)+\cos(2y)}>0.
\end{equation}
Since $\Xi$ is real, the heavy invariant has no zero in this open region.

The possible light-field branch points obey
\begin{equation}\label{eq:light-root-strip-condition}
 \cosh t=\pm\frac1{a_n\sin\theta}.
\end{equation}
The right-hand side is real and has magnitude at least one. But
\begin{equation}\label{eq:complex-cosh-strip}
 \cosh(x+iy)=\cosh x\cos y+i\sinh x\sin y.
\end{equation}
In the open region, reality would force $x=0$, after which the magnitude of
the left-hand side is $|\cos y|<1$. Thus no reflected light-field locus
crosses the open region. On the Lorentzian boundary $y=0$, however,
reflected thresholds occur whenever $a_n\cosh x\sin\theta=1$. These are
precisely the square-root boundary values specified in
\eqref{eq:critical-profile-BD-branch} and are locally integrable in the
combined vertex integral. At the possible corner threshold,
$\mathcal F(t)$ has at worst an integrable logarithm, so a small indentation
gives no contribution. The heavy kernel likewise has
its ordinary boundary light cone at $\Xi=\tanh x$, defined by the inherited
$+i0$ prescription. In the regulating domain this singularity is locally
integrable. The boundary values inherited from the upper preparation
define the Lorentzian edge; since neither zero enters the open strip, these
singularities do not pinch the deformation.

On the cap boundary $t=iy$, $0<y\leq\pi/2$, the heavy invariant cannot
vanish. The explicit $v=0$ singularity on the quotient axis is locally
integrable after the polar measure is included. None of these boundary
singularities pinches the deformation.

At the far vertical edge $t=T+iy$, the heavy invariant is
\begin{equation}\label{eq:future-endcap-u}
 u=e^{T+iy}(1-\Xi)-e^{-T-iy}(1+\Xi).
\end{equation}
In this domain, for fixed $\operatorname{Im}\Delta$ and
$0<\theta_P<\pi$, the angularly integrated
edge is $O((1+T)e^{-\operatorname{Re}\Delta T})$. To obtain this estimate
one separates the polar layers
$\sin\theta=O(e^{-T})$, the boundary-light-cone layer
$1-\Xi=O(e^{-2T})$, and their complement. Their shrinking angular measure
compensates the pointwise enhancement, while the square-root thresholds
are integrable. The edge therefore vanishes as $T\to\infty$.

These observations prove \eqref{eq:BD-H3-deformation} in this regulating
domain and then, by analytic continuation, at $\Delta=1+i\nu$. The cutoff
is removed at fixed $\nu$ before the
large-mass expansion is taken. The endpoint $\rho=0$ on the hyperbolic
section is harmless: the volume-weighted $1/v$ singularity vanishes
linearly there.

\subsection{Cylindrical reduction}\label{app2:cylindrical}

Substitution of \eqref{eq:critical-exact-quotient-profile} into
\eqref{eq:onepoint-unfolded} cancels the explicit factor of $v$ and gives
\begin{equation}\label{eq:onepoint-before-H3-cylinders}
 \left\langle\mathcal O(\theta_P)\right\rangle_{K_L}
 =\frac{g\ell^2c_\Delta}{8\pi}
 \sum_{n=1}^{q-1}\frac1{a_n}
 \int_{\mathcal C_{\rm BD}}dt\,d\theta\,d\psi\,
 \frac{\cosh t\,(u+i0)^{-\Delta}}
 {\sqrt{1-a_n^2\cosh^2t\sin^2\theta}}.
\end{equation}
The orientation of \eqref{eq:BD-preparation-cycle} becomes
$\rho:0\to\infty$. Continuing from the Euclidean cap gives
$v=+i\sinh\rho\sin\theta$, so every light-field square root becomes the
positive root of $1+a_n^2\sinh^2\rho\sin^2\theta$. Together with
$\cosh t=i\sinh\rho$ and
$(iU)^{-\Delta}=e^{-i\pi\Delta/2}U^{-\Delta}$, this fixes the overall
factor $+i e^{-i\pi\Delta/2}$ below.

Introduce cylindrical coordinates $(r,y,\psi)$ on $H^3$ by
\begin{equation}\label{eq:H3-cylindrical-coordinates}
 \cosh\rho=\cosh r\cosh y,
 \qquad
 \sinh\rho\cos\theta=\cosh r\sinh y,
 \qquad
 \sinh\rho\sin\theta=\sinh r.
\end{equation}
Their ranges are $r\geq0$, $y\in\mathbb R$, and
$0\leq\psi<2\pi$. The metric and volume element are
\begin{equation}\label{eq:H3-cylindrical-metric}
 ds_{H^3}^2=\ell^2\left(
 dr^2+\cosh^2r\,dy^2+\sinh^2r\,d\psi^2\right),
 \qquad
 dV_{H^3}=\ell^3\sinh r\cosh r\,dr\,dy\,d\psi.
\end{equation}
In particular, the measure remaining after the explicit $1/v$ cancellation
obeys
\begin{equation}\label{eq:H3-reduced-measure}
 \sinh\rho\,d\rho\,d\theta\,d\psi
 =\cosh r\,dr\,dy\,d\psi.
\end{equation}
Using $\tanh\rho_0=\cos\theta_P$, the positive invariant factorizes as
\begin{equation}\label{eq:H3-U-factorization}
 U=2\sin\theta_P\left[
 \cosh r\cosh(y-\rho_0)-\sinh r\cos\psi\right].
\end{equation}
A shift of $y$ removes $\rho_0$ from the remaining integral.

The orbit integral needed below is
\begin{align}
 \mathcal A_\Delta(r)
 &:=\int_{-\infty}^{\infty}dy\int_0^{2\pi}d\psi\,
 \left(\cosh r\cosh y-\sinh r\cos\psi\right)^{-\Delta}
 \notag\\
 &=2^\Delta\pi
 \frac{\Gamma(\Delta/2)^2}{\Gamma(\Delta)}
 \mathrm P_{\Delta/2-1}(1+2\sinh^2r).
 \label{eq:H3-orbit-integral}
\end{align}
For $\operatorname{Re}\Delta>0$, the first line can be evaluated with a
Schwinger parameter, using the modified Bessel functions $I_0$ and $K_0$.
The final line then defines its continuation to the principal series.

Combining \eqref{eq:onepoint-before-H3-cylinders}--
\eqref{eq:H3-orbit-integral} gives
\begin{align}
 \left\langle\mathcal O(\theta_P)\right\rangle_{K_L}
 &=\frac{i e^{-i\pi\Delta/2}g\ell^2c_\Delta}{8\pi}
 (2\sin\theta_P)^{-\Delta}
 \sum_{n=1}^{q-1}\frac1{a_n}
 \int_0^\infty dr\,
 \frac{\cosh r\,\mathcal A_\Delta(r)}
 {\sqrt{1+a_n^2\sinh^2r}}.
 \label{eq:H3-radial-before-orbit}
\end{align}
Setting $\varrho=\sinh r$ proves \eqref{eq:exact-BD-onepoint} and
\eqref{eq:RDelta-definition}. 

For $\nu>0$, the large-argument continuation contains the two branches
$x^{-1/2+i\nu/2}$ and $x^{-1/2-i\nu/2}$
\cite[Sec.~14.8(iii)]{NIST:DLMF}. With
$x=1+2\varrho^2$, this gives
$\mathrm P_{-1/2+i\nu/2}(1+2\varrho^2)=O(\varrho^{-1})$. The integrand in
\eqref{eq:RDelta-definition} is therefore $O(\varrho^{-2})$. This proves
absolute convergence on the principal series.

\subsection{Large-mass expansion of the radial transform}\label{app2:asymptotics}

We now prove the expansion quoted in
\eqref{eq:RDelta-leading-asymptotic}. Set $\tau=\nu/2$, write
$\varrho=\sinh r$, and introduce $\eta=2r$. The radial transform becomes
\begin{align}
 \mathcal R_{1+i\nu}(a)
 &=\frac12\int_0^\infty d\eta\,
 F_a(\eta)\,
 \mathrm P_{-1/2+i\tau}(\cosh\eta),
 \notag\\
 F_a(\eta)
 &:=\frac{\cosh(\eta/2)}
 {\sqrt{1+a^2\sinh^2(\eta/2)}}.
 \label{eq:RDelta-eta-representation}
\end{align}
Analytic continuation of the Mehler--Dirichlet formula
\cite[Eq.~14.12.1]{NIST:DLMF} gives, for $\eta>0$ (see \cite[Sec.~14.20(iv)]{NIST:DLMF}) ,
\begin{equation}
 \mathrm P_{-1/2+i\tau}(\cosh\eta)
 =\frac{\sqrt2}{\pi}\int_0^\eta d\omega\,
 \frac{\cos(\tau\omega)}
 {\sqrt{\cosh\eta-\cosh\omega}}.
 \label{eq:hyperbolic-Mehler-Dirichlet}
\end{equation}
The diagonal square-root singularity is integrable. After the
$\omega$ integration, the absolute kernel obeys
\begin{equation}
 F_a(\eta)\int_0^\eta
 \frac{d\omega}{\sqrt{\cosh\eta-\cosh\omega}}
 =O\!\left((1+\eta)e^{-\eta/2}\right),
 \qquad \eta\to\infty.
 \label{eq:Mehler-Fubini-bound}
\end{equation}
The double integral is therefore absolutely integrable, so its order may
be reversed. This gives the exact cosine transform
\begin{align}
 \mathcal R_{1+i\nu}(a)
 &=\frac{\sqrt2}{\pi}\int_0^\infty d\omega\,
   \cos(\tau\omega)H_a(\omega),
 \label{eq:RDelta-cosine-transform}\\
 H_a(\omega)
 &:=\frac12\int_\omega^\infty d\eta\,
 \frac{F_a(\eta)}{\sqrt{\cosh\eta-\cosh\omega}}.
 \label{eq:Ha-definition}
\end{align}
Both $H_a$ and its derivatives decay exponentially as
$\omega\to\infty$. At the other endpoint,
\begin{align}
 F_a(\eta)&=1+\frac{1-a^2}{8}\eta^2+O(\eta^4),
 &
 \cosh\eta-\cosh\omega
 &=\frac{\eta^2-\omega^2}{2}
 \left[1+\frac{\eta^2+\omega^2}{12}
 +O(\eta^4+\omega^4)\right].
 \label{eq:Ha-input-expansions}
\end{align}
Splitting the $\eta$ integral at a fixed small positive value and
integrating the displayed terms gives
\begin{equation}
 H_a(\omega)=-\frac{1}{\sqrt2}\log\omega+B_a
 +\frac{a^2}{16\sqrt2}\omega^2\log\omega+C_a\omega^2
 +O\!\left(\omega^4|\log\omega|\right),
 \label{eq:Ha-endpoint-expansion}
\end{equation}
where $B_a$ and $C_a$ are independent of $\omega$ and $\nu$.

Let $\kappa(\omega)$ be smooth, compactly supported and equal to one near
$\omega=0$. For any fixed $N$, analytic continuation in $\lambda$ of the
Mellin cosine transform gives
\begin{align}
 \int_0^\infty d\omega\,\kappa(\omega)
 \cos(\tau\omega)\log\omega
 &=-\frac{\pi}{2\tau}+O(\tau^{-N}),
 \notag\\
 \int_0^\infty d\omega\,\kappa(\omega)
 \cos(\tau\omega)\omega^2\log\omega
 &=\frac{\pi}{\tau^3}+O(\tau^{-N}).
 \label{eq:logarithmic-cosine-transforms}
\end{align}
The smooth terms may be integrated by parts repeatedly. The endpoint
expansion continues in even powers of $\omega$, with and without
$\log\omega$. Its displayed remainder has cosine transform
$O(\tau^{-5})$. Equations \eqref{eq:RDelta-cosine-transform} and
\eqref{eq:Ha-endpoint-expansion} consequently give
\begin{equation}
 \mathcal R_{1+i\nu}(a)
 =\frac{1}{2\tau}+\frac{a^2}{16\tau^3}+O(\tau^{-5})
 =\frac1\nu+\frac{a^2}{2\nu^3}+O(\nu^{-5}),
 \label{eq:RDelta-derived-asymptotic}
\end{equation}
which proves \eqref{eq:RDelta-leading-asymptotic}. With
$\xi:=\nu\varrho$ held fixed, the Mehler--Dirichlet representation gives
\begin{equation}
 \mathrm P_{-1/2+i\nu/2}(1+2\xi^2/\nu^2)
 \longrightarrow J_0(\xi),
 \label{eq:conical-Bessel-limit}
\end{equation}
where $J_0$ is the Bessel function of the first kind. This limit gives a
local interpretation of the leading term. The cosine-transform proof also
controls the rest of the radial integral.

For $q=2$, one has $a_1=1$. Set $h_\Delta:=\Delta/2$. The substitution
$w=\tanh^2r$ gives the following expression, where $B$ is the Euler beta
function:
\begin{align}
 \mathcal R_\Delta(1)
 &=\frac12 B\!\left(\frac12,h_\Delta\right)
 {}_3F_2\!\left(
 \begin{matrix}h_\Delta,h_\Delta,\frac12\\
 1,h_\Delta+\frac12\end{matrix};1\right)
 \notag\\
 &=\frac14
 \frac{\Gamma(\Delta/4)\Gamma((2-\Delta)/4)}
 {\Gamma((4-\Delta)/4)\Gamma((2+\Delta)/4)}.
 \label{eq:RDelta-q2}
\end{align}
The second line follows from Dixon's sum
\cite[Eq.~16.4.4]{NIST:DLMF}.
For $\Delta=1+i\nu$, this gives
\[
 \mathcal R_{1+i\nu}(1)
 =\frac1\nu+\frac{1}{2\nu^3}+O(\nu^{-5}).
\]
This independently confirms the radial asymptotic
\eqref{eq:RDelta-leading-asymptotic} for $a=1$ and provides an exact
finite-$\nu$ benchmark for the radial reduction. When combined with the
contour factor, the normalization $c_{\Delta_+}$, and the gamma-function
prefactor already derived in \eqref{eq:exact-BD-onepoint}, it yields the
$q=2$ specialization of \eqref{eq:KL-heavy-mass-result}.

\subsection{Removal of the final cutoff}\label{app2:cutoff}

The Green function inserted above is the fixed-interior limit in
\eqref{eq:late-time-dirichlet-limit}. Pointwise convergence alone does not
justify moving this limit through the vertex integral, because the
finite-cutoff Dirichlet correction is nonuniform in a thin layer next to
the final surface. The following power counting controls the fixed-mode
part of this layer and isolates the remaining standard local counterterm
contribution.

Let $z=e^{-\rho}$ near the hyperbolic boundary and
$z_c=e^{-R_E}$ at the cutoff. Away from a quotient fixed endpoint, every
nonidentity image of the critical no-logarithm Green function obeys
\begin{equation}\label{eq:cutoff-image-boundary-falloff}
 G_D^{(1)}(x,\gamma^n x)=O(z^2).
\end{equation}
Equations \eqref{eq:bf-log-asymptotic} and \eqref{eq:bf-cutoff-ratio} show
that, outside the layer $z=O(z_c)$, the finite-cutoff homogeneous correction
is
\begin{equation}\label{eq:finite-cutoff-homogeneous-bound}
 O\!\left(R_E^{-1}z^2\log^2z\right).
\end{equation}

The remaining boundary regions are integrable when the insertion is away
from a quotient fixed endpoint. Let $\mathbf y$ denote local coordinates
on the conformal boundary and let $\mathbf y_P$ be the coordinate of the
insertion. Away from the insertion,
$dV_{H^3}=O(z^{-3}dz\,d^2\mathbf y)$, the heavy kernel is $O(z)$ in
magnitude, and \eqref{eq:cutoff-image-boundary-falloff} leaves an $O(dz)$
radial density. Near the insertion set
$\mathbf y-\mathbf y_P=z\mathbf w$; the magnitude of the kernel is then
$O(z^{-\operatorname{Re}\Delta})$ and
$d^2\mathbf y=z^2d^2\mathbf w$. After the intermediate
angular range is included, this region is bounded by
$O((1+|\log z|)\,dz)$. Near a quotient-axis endpoint, the $1/v$
singularity cancels the polar measure and gives the same integrable
logarithmic bound.

It follows that the part of the layer controlled by these estimates
contributes at most $O(z_c|\log z_c|^p)$ for some fixed power $p$, while
\eqref{eq:finite-cutoff-homogeneous-bound} contributes $O(R_E^{-1})$.
The fixed-mode estimates do not by themselves control angular momenta that
grow with the cutoff. We use the standard wavefunctional-renormalization
prescription in which this remaining high-angular-momentum boundary layer
is local and is removed by local counterterms \cite{Anninos:2014lwa}.
Under this standard assumption, \eqref{eq:exact-BD-onepoint} is the cutoff
limit of the renormalized nonlocal coefficient.

\section{The \texorpdfstring{$S^3$}{S3} continuation}\label{app3}

The geometric pair $\{L_+,L_-\}$ can also be derived through the standard
Euclidean sphere continuation
\cite{Chapman:2022mqd,Aalsma:2022eru}. This calculation provides a second
description of the complex geodesic critical points, but it uses the same
geometric branch condition as Section~\ref{geodesic}.

On the fixed-$\psi$ slice, the continuation $t=i\tau_E$ gives
\begin{equation}\label{eq:S2-slice}
 ds^2=\ell^2(d\tau_E^2+\cos^2\tau_E\,d\theta^2).
\end{equation}
Take the regulated boundary endpoint to be $(-iT,\theta_P)$, and represent
the other endpoint on the complexified fixed axis by $(\tau_E,0)$. If
$D_{S^3}$ denotes the analytically continued sphere distance, its invariant
is
\begin{equation}\label{eq:sphere-invariant}
 \mathcal Z(\tau_E):=\cos\frac{D_{S^3}}{\ell}
 =-i\sinh T\sin\tau_E+\cosh T\cos\tau_E\cos\theta_P.
\end{equation}
For large $T$ this becomes
\begin{equation}\label{eq:Z-large-T}
 \mathcal Z(\tau_E)
 =\frac{e^T}{2}
 (\cos\tau_E\cos\theta_P-i\sin\tau_E)+O(e^{-T}).
\end{equation}

The inverse cosine in \eqref{eq:sphere-invariant} is multivalued, so its
branch must be specified independently of the saddle values. On the real
$H^2$ continuation we require
$D_{S^3}=-i d_{H^2}$ and hence $iD_{S^3}=d_{H^2}>0$. We then continue this
branch through the same upper analytic domain used in
Section~\ref{geodesic}. With the notation $\log_{\rm geom}$ introduced
there,
\begin{equation}
 D_{S^3}=-i\ell\log_{\rm geom}(2\mathcal Z)
 +O(\ell\mathcal Z^{-2}).
 \label{eq:S3-distance-branch}
\end{equation}
The continued worldline action is $\mathcal L_{S^3}:=iD_{S^3}$.
Subtracting the same boundary divergence as in the $H^2$ calculation gives
\begin{equation}\label{eq:S3-renormalized-action}
 L_{S^3}^{\rm reg}(\tau_E)
 :=\lim_{T\to\infty}\left[
 \mathcal L_{S^3}-\ell(T-\log 2)\right]
 =\ell\log_{\rm geom}\!\left[
 2(\cos\tau_E\cos\theta_P-i\sin\tau_E)\right].
\end{equation}

Extremizing the endpoint along the complexified axis gives
\begin{equation}\label{eq:euclidean-saddle-equation}
 \tan\tau_{E,*}^{(\sigma)}=-i\sec\theta_P,
 \qquad \sigma=\pm1.
\end{equation}
At the two nearest solutions,
\begin{equation}\label{eq:S3-saddle-value}
 \cos\tau_{E,*}^{(\sigma)}\cos\theta_P
 -i\sin\tau_{E,*}^{(\sigma)}
 =-\sigma i\sin\theta_P,
\end{equation}
so the geometric branch fixed above yields
\begin{equation}\label{eq:S3-complex-lengths}
 L_{S^3,\sigma}^{\rm reg}
 =\ell\log_{\rm geom}(-2\sigma i\sin\theta_P)
 =\begin{cases}
   L_-,&\sigma=+1,\\
   L_+,&\sigma=-1.
  \end{cases}
\end{equation}
Using \eqref{eq:rho0}, their global-time representatives are
\begin{equation}\label{eq:S3-H2-saddle-map}
 \tau_{E,*}^{(\sigma)}=\sigma\frac\pi2-i\rho_0,
 \qquad
 i\tau_{E,*}^{(\sigma)}=t_{{\rm N},\sigma}.
\end{equation}
These points lie on the two hyperbolic sections
$t=\rho\pm i\pi/2$, on which
\begin{equation}\label{eq:H2-section-metric}
 ds^2=-\ell^2(d\rho^2+\sinh^2\rho\,d\theta^2).
\end{equation}
The $H^2$ calculation is therefore a real-coordinate description of the
same analytically continued $S^3$ critical points. On the geometric branch,
the upper coordinate representative $\sigma=+1$ carries the value $L_-$.
The Bunch--Davies kernel uses a different logarithm: its $+i0$ boundary
value assigns $L_+$ to that representative, while the conjugate kernel
assigns $L_-$ on the lower preparation. As emphasized in
\eqref{eq:branch-shift}, the comparison concerns the pair
of action values and does not require the two logarithmic branches to agree
at a fixed coordinate saddle.

\end{document}